\documentclass[showpacs,showkeys,preprintnumbers,amsmath,amssymb,10pt,prd]{llncs}
\usepackage{graphicx}
\usepackage{hyperref}
\usepackage{bm}
\usepackage{dcolumn}
\usepackage[mathscr]{eucal}
\usepackage{color}
\usepackage{pst-plot}
\usepackage{amsmath}
\usepackage{comment}
\definecolor{mygray}{gray}{0.4}

\makeindex

\begin{document}
\title{The Computing Spacetime}
\author{Fotini Markopoulou}

\institute{Perimeter Institute for Theoretical Physics, 
Waterloo, Ontario N2L 2Y5
Canada
\\
and
\\
 University of Waterloo, Waterloo, Ontario N2L 3G1, Canada 
\\
and
\\
Max Planck Institute for Gravitational Physics, Albert Einstein Institute,\\
Am M\"uhlenberg 1, Golm, D-14476 Golm, Germany
\\
\email{fotinimk@gmail.com}
}

\maketitle


\begin{abstract}
The idea that the Universe is a program in a giant quantum computer is both fascinating and suffers from various problems.  Nonetheless, it can provide a unified picture of physics and this can be very useful for the problem of Quantum Gravity where such a unification is necessary.   In previous work we proposed Quantum Graphity, a simple way to model a dynamical spacetime  as a quantum computation.  In this paper, we give an easily readable introduction to the idea of the universe as a quantum computation, the problem of quantum gravity, and the graphity models. 
\end{abstract}

\section{Introduction}

That the Universe can be thought of as a giant computation is a straightforward corollary of the existence of a universal Turing machine.   The basic idea (nicely summarized, for example, by Deutsch in \cite{Deut}) goes as follows. 
The laws of physics allow for a machine, the universal Turing machine, such that its possible motions correspond to all possible motions of all possible physical objects.  That is, a universal quantum computer can simulate every physical entity and its behavior.  This means that physics, the study of all possible physical systems, is isomorphic to the study of all programs that could run  on a universal quantum computer.  In short, our universe can be thought of as software running on a universal computer.

Should this logical inference affect our understanding of physics, or even change the way we do science?  Several different lines of thought say yes, an idea most concretely articulated in the field of cellular automata and quantum information theory.  

In 1969, Konrad Zuse, in his book {\em Calculating Space}, proposed that the physical laws of the universe are fundamentally discrete, and that the entire universe is the output of a deterministic computation on a giant cellular automaton \cite{Zuse}.
Cellular automata (CA) are regular grid of cells, and  each cell can be in one of a finite number of states, usually on or/off, or black/white.  An initial state of the CA is updated in global discrete time steps, in which each cell's new state changes as a function of its old state and that of a small number of neighbors.  
 A concrete example of Zuse's vision is  Conway's  Game of Life.  The rules are simple: If a cell has 2 black neighbors, it stays the same; if it has 3 black neighbors, it becomes black; otherwise it becomes white. The result is remarkably rich behavior on the border between randomness and order. A striking feature is the occurrence of gliders, small groups of cells that appear to move like independent emergent entities. It is possible to arrange the automaton so that the gliders interact to perform computations, and it can be shown that the Game of Life is a universal Turing machine \cite{GLTu}. 
It is simple to see how this  evokes the possibility that we live in a giant CA \cite{Den}: 
In our CA Universe, what we think of as elementary particles may just be emergent gliders. Since CAs exist that are Turing machines, it is in principle possible to have any kind of glider behavior generated by a CA, including gliders observing the laws of elementary particle physics.  We don't know how these are generated because we have no access to the microscopic cells, so we make physical theories about particle-like objects, but, in reality, we live in a CA.

Quantum information theory has given a new and interesting twist on the Universe as a Computation.  A common idea that is advocated by many practitioners in this field is that everything fundamentally is information, an old idea that can be traced at least back to Wheeler's influential {\em it from bit} \cite{Whee}.   In that view, all interactions between physical systems in the universe  are instances of information processing, and the information involved in those processes is more primary than the physical systems themselves. Instead of thinking of particles as colliding, we should think of the information content of the particles being involved in a computation.  By simple interpolation, the entire universe is nothing but a giant computation. As Lloyd puts it in \cite{Lloy}, the universe computes ``its own dynamical evolution; as the computation proceeds, reality unfolds''.

These are fascinating ideas when loosely interpreted, but with obvious problems, including: 
1. What does it mean that information is more fundamental than its physical instantiation?  
2. Since any observation we can make, and anything physics describes, is just the program, there is no way to know the hardware that runs that program.  The program can perhaps give us some hints as to what machine could efficiently run it, but at the end of the day this scenario assumes a fundamentally unknowable machine. 
3. Is that machine running just one program, our universe?  If yes,  how is that ``mother computation'' chosen?  If no, we need a  meta-program that runs multiple programs, a computer version of the {\em multiverse} idea \cite{Tegmark}. By one more iteration, multiple computers, each running multiple programs, are a logical possibility, leading to an extreme form of a multiverse.  Or are we secretly assuming a Programmer?  4. The idea requires that all of physics is computable.  
5. The CA Universe, in addition, advocates that the universe is fundamentally discrete.  
  Fundamental discreteness is a very old and attractive idea but it remains to be seen whether it can be reconciled with observable physics, and, in particular, with  quantum mechanics and Lorentz invariance.  Quantum mechanics makes essential use of the complex numbers,  a continuous field.  
 It is, of course, logically possible to push fundamental discreteness to an extremely small scale, perhaps the Planck scale, and claim that the world appears continuous only by approximation, because we have no access to that small scale.  This is where Lorentz invariance comes in.  The Lorentz transformations form a non-compact group,  meaning that by boosting an observer sufficiently, we can blow up any tiny amount of discreteness to arbitrarily large size. Depending on the details of the physics, even scales smaller than the Planck scale  can thus become observationally accessible.  Reconciling observational constraints on Lorentz invariance violations and fundamental discreteness  is a very active subject of research in quantum gravity and quantum gravity phenomenology \cite{Lorentz}.  

At the end of the day, the Universe as a Computation idea
 may just reflect the current way we understand and bring order to our surroundings.  It is 
possible that all it does is shift us a little  from the ``Blind Watchmaker'' to the ``Blind Programmer''.  
I find it very likely that the
 Universe as Computation is a 
 culturally determined and temporary idea.  In any case, fun as it may be to speculate about the universe being a computer, it is rather sterile to do so in the abstract.  The interesting  question is whether this scenario can  be put to good uses: Does it give us useful new tools and methods with which we can solve problems we couldn't solve before? Does it raise new interesting questions?  
The purpose of this article is to argue that it does, and that the relevant area of physics to explore and use the idea of the Universe as a Computation is the field of Quantum Gravity and Quantum Cosmology.  

 If there is merit to this idea, it should be useful in the physics of the entire universe.  This is the research field of Quantum Gravity and Quantum Cosmology.  
Quantum Gravity needs to unify quantum field theory, the physics of matter, with general relativity, the physics of spacetime, into a single consistent theory.  
The universe as a Computation suggests a new kind of unification: physical systems and their dynamics can be represented in terms of their information content and their dynamics is the processing of that information. 

We will illustrate this view with an example.  
In \cite{KMS,HMLCSM}, we initiated a study of quantum gravity using spin systems as toy models for emergent geometry and gravity. These models, which we named {\em quantum graphity models}, are based on quantum networks with no a priori geometric notions.  We have repeatedly found the quantum information perspective to be useful, both as a tool chest (for example, as we will see, the Lieb-Robinson speed of information propagation can be used to derive the speed of light \cite{LR}, or error correction to define conserved quantities \cite{braids}) and as an aid to conceptual clarity:  the information theoretic language allows us to do physics without reference to a background geometry. 

The purpose of this introductory article is to illustrate these ideas in a brief and self-contained format and invite discussion and exchange of ideas between the fields of quantum gravity and computer science.  Technical details the reader can find in the suggested references.  In the next Section, we state the problem of Quantum Gravity in terms of the breakdown of classical spacetime at Planck scale and the problem of time.  In Section 3, we summarize the basics of Quantum Graphity, the representation of pre-geometry as qubits of adjacency, an example of interacting matter-geometry model, a sketch of the derivation of the speed of light from the fundamental dynamics, the toy trapped surfaces that arise in these systems, and the mechanism by which matter inside that world sees an emergent curved geometry.  We briefly summarize our conclusions in Section 4.

\section{The Problem of Quantum Gravity}

The field of Quantum Gravity is the attempt  to unify General Relativity and Quantum Theory.    In spite of their impressive successes, the two theories leave us with a gap in situations in which the quantum effects of the gravitational field become important.  This hampers our understanding of some of  the most fascinating modern physics, such as the physics of black holes \cite{Susk},  Hawking radiation \cite{JacBH}, and the very early universe \cite{cosmo}, or leads to absurdities such as the cosmological constant problem \cite{CCP}. 
These are all 
situations in which the  curvature of spacetime is so high that  we are not confident in the reliability of quantum field theory calculations. 

The length scale we expect quantum gravitational effects to become significant is, by dimensional analysis, the Planck length, $l_{\rm Pl}=\sqrt\frac{G_N\hbar}{c^3}$, the combination of Newton's constant $G_N$, Planck's constant $\hbar$, and the speed of light $c$.   This is incredibly small, $l_{\rm Pl}\sim 10^{-35}m$, or, equivalently, corresponds to energy scales of $10^{19}GeV$.  
At Planckian scales, the concepts of size and distance break down.  Any microscopic probe energetic enough to precisely measure a Planck-sized object needs to be so energetic (to measure $l_{Pl}$, its Compton wavelength must be $\sim l_{Pl}$) that it would
completely distort the region of space it was supposed to measure.   In this sense, the notion of a classical spacetime manifold breaks down at Planck scale.   A quantum theory of gravity that reconciles general relativity and quantum theory, or replaces them, is required to understand physics, including spacetime, at that scale.  

In spite of decades of research, finding a satisfactory quantum theory of gravity still eludes us.  
Much of the difficulty in reconciling general relativity and quantum theory comes from the fundamentally different assumptions that these theories make on how the universe works.   General relativity describes spacetime as a manifold ${\cal M}$ with a dynamical metric field $g_{\mu\nu}$, and gravity as the curvature that spacetime. Quantum field theory describes particle fields on a flat and  fixed spacetime. Naive quantization of gravity, treating it as another quantum field, leads to nonsense as gravity is non-renormalizable.  
The difference between the two theories can be phrased in terms of the way each treats {\em time}.
A fundamental lesson of general relativity is that there is no fixed spacetime background spacetime: {\em geometry tells matter where to go and matter tells geometry how to curve}. The spacetime geometry is a dynamical field.  In addition, physical quantities are invariant under diffeomorphisms of ${\cal M}$.  This means (roughly) that 
 general relativity is a relational theory, i.e.,  the only physically relevant information is the relationship between different events in spacetime \cite{Stac}.
On the other hand, quantum theory requires a fixed background spacetime, 
either a Newtonian one (quantum mechanics), or a fixed Minkowski spacetime (quantum field theory).  Time in quantum theory is not a dynamical field, it is a background parameter.  

Turning the spacetime geometry into a quantum field is possible and the task of conservative approaches to quantum gravity such as Loop Quantum Gravity \cite{LQG}.  The result, however, of such quantizations is peculiar.  We obtain a so-called {\em wavefunction of the universe} $|\Psi_U\rangle$, i.e., the diffeomorphism invariant quantization of the metric $g_{\mu\nu}$ projected on a spatial slice of ${\cal M}$. Instead of a Schr\"odiger equation, the evolution of $|\Psi_U\rangle$ is governed by the {\em Wheeler-deWitt equation}:
\begin{equation}
\widehat H |\Psi_U\rangle=0,
\end{equation}
where $\widehat H$ is the quantization of the ``projection'' of the Einstein equations in the direction normal to the spatial slice (for the actual details of this procedure, see, for example, \cite{Rovelli}).  
The Wheeler-deWitt equation is peculiar on two (related) counts: it describes the evolution of the entire universe, not just a localized system as in the Schr\"odinger equation, and the right hand side is zero (not time evolution).  That zero can be traced to the diffeomorphism invariance of general relativity and the fact that the Einstein equations describe the dynamics of the {\em entire} universe.  The diffeomorphism symmetry gets mixed up with evolution in ways that are very difficult to untangle\footnote{
For a classic review of the longstanding effort to find gravity's true degrees of freedom (metrics modulo diffeomorphisms) see \cite{Isha}.
}\footnote{
$|\Psi_U\rangle$ is also where the subject of quantum cosmology comes in.  General relativity is a {\em cosmological theory}, meaning that it describes the entire universe.  Making  this quantum raises numerous issues with the standard interpretation of quantum mechanics, such as the role of the observer and emergence of classicality.  Such issues are the subject of quantum cosmology.}.

Despite repeated attacks on the problem from multiple fronts, finding a satisfactory quantum theory of gravity remains an open problem.  Much more can be said about this, but the purpose of the present note is to point out that, since quantum gravity needs to unify quantum theory and general relativity, a unification of the corresponding  descriptions of the physical world is required, and that quantum information theory can provide this.  Reducing both  quantum fields and differential manifolds
to their information theoretic content can provide a common framework.  The  Universe 
as a Computation can, in that sense, be seen as a useful and practical tool to solve a long-standing problem.  Note that do not need to resolve whether information precedes its physical instantiation, or answer most of the problems listed above in order to put this notion to useful work.  All we need is that an information theoretic description is possible, both for the physics of matter and for the physics of space-time.  We have been pursuing this idea in the {\em Quantum Graphity models} for quantum gravity and we will give a concrete example of such a model in the next Section.

\section{Quantum Graphity}

Quantum Graphity models \cite{KMS,HMLCSM} are spin system toy models for emergent geometry and gravity.  They are based on  graphs whose adjacency is quantum  and dynamical: their edges  can be on (connected), off (disconnected),
or in a superposition of on and off. We can  interpret
the graph as pregeometry (the connectivity of
the graph tells us who is neighbouring whom). A
particular graphity model is given by such graph states
evolving under a local Hamiltonian. The graphity model of \cite{HMLCSM}, for example, which will describe in the rest of this section, is a toy model for interacting matter and geometry,  a Bose-Hubbard model where the interactions, or adjacencies, are quantum variables.

\subsection{Qubits of adjacency}

Let us  assume a universe consisting of $N$ fundamental constituents, 
 systems labeled by $i=1,É,N$.  These are quantum mechanical, so we have  $\{ \mathscr H_i\}; i=1,...,N$ Hibert spaces.
 Let $K_N$ denote the complete graph that has these $N$ systems as its vertices, a graph with $\frac{N(N-1)}{2}$ links ${\bf e}\equiv(i,j)$.  To every such  ${\bf e}$   we 
associate a Hilbert space $\mathscr H_{\bf e}\simeq {\bf C}^2$ .  Basis states on $\mathscr H_{\bf e}$ can be labeled by $|1\rangle, |0\rangle$, and we choose to interpret  $|1\rangle$ as the link ${\bf e}$ being {\em on}, or present, and  $|0\rangle$ as the link being {\em off}, or missing.  The total Hilbert space corresponding to $K_N$ then is ${\cal H}_{graph}=\bigotimes_{{\bf e}=1}^{N(N-1)/2}{\cal H}_{\bf e}$.  

Our choice of basis in ${\cal H}_{\bf e}$ means that every basis state in ${\cal H}_{graph}$ corresponds to a subgraph of $K_N$.  A generic state $|\Psi_{graph}\rangle\in{\cal H}_{graph}$ is a quantum superposition of subgraphs of $K_N$.  For $N$ very large, the state space contains superpositions of all possible finite graphs.  
By analogy with the adjacency matrix of a graph, we call $\mathscr H_{\bf e}$ a {\em qubit of adjacency}.  States in ${\cal H}_{graph}$ then provide a simple discrete precursor to quantum geometry.  Note, however, that since we cannot assume a pre-existing spacetime on which our $N$ systems live, we cannot interpret the $N$ vertices of $K_N$ as points in that spacetime.  That is, we do not have a discretization of a geometry, the geometry corresponding to a state is to be {\em inferred} from the behavior of matter interacting with $|\Psi_{graph}\rangle$.  

To see how this works, let us next define a simple form of matter.

\subsection{Interacting matter and geometry}

We will assign simple matter degrees of freedom to the vertices of $K_N$ by assigning 
 the Hilbert space  $\mathscr H_i$ of a harmonic oscillator to each vertex $i$. We denote its creation and annihilation 
operators by $\hat b^\dagger_i,\hat b_i$ respectively, where $\hat b^\dagger_i|0\rangle_i=|1\rangle_i,\hat b_i|1\rangle_i=|0\rangle_i$,  satisfying the usual bosonic relations, $[\hat b^\dagger, \hat b^\dagger]=0=[\hat b,\hat b], [\hat b, \hat b^\dagger]=1$. Our $N$ physical systems then 
are $N$ bosonic particles and the total Hilbert space for these bosons is given by
$
\mathscr H_{bosons} = \bigotimes_{i=1}^N \mathscr H_i.
$

The total Hilbert space of the theory is the state space of the combined matter and connectivity degrees of freedom, 
$
{\mathscr H} = \mathscr H_{bosons}\otimes \mathscr H_{graph}.$ 
A basis state in ${\cal H}$ has the form
$
|\Psi\rangle \equiv |\Psi_{bosons}\rangle\otimes|\Psi_{graph}\rangle
 \equiv |n_1,...,n_N\rangle\otimes |e_1,...,e_{\frac{N(N-1)}{2}}\rangle.
$
The first factor tells us how many bosons there are at every site $i$, while the second factor tells us which pairs ${\bf e}$ interact.  This is an unusual bosonic system, as  
the structure of interactions is now promoted to a quantum degree of freedom.

 This is interesting as generic state can  be a quantum 
superposition of ``interactions''. For example, consider the systems $i$ and $j$ in the state
$
|\phi_{ij}\rangle = (|10\rangle\otimes|1\rangle_{ij}+ |10\rangle\otimes|0\rangle_{ij})/{\sqrt{2}}$. 
This state describes a particle in $i$ and no particle in $j$, and a 
quantum superposition between $i$ and $j$ interacting or not. The state,
$
|\phi_{ij}\rangle = (|00\rangle\otimes|1\rangle_{ij}+ |11\rangle\otimes|0\rangle_{ij})/{\sqrt{2}},
$
represents a different superposition, in which the bosonic degrees of freedom and the graph degrees of freedom are 
entangled. It is a significant feature of the model that {matter can be entangled with geometry}.

In \cite{HMLCSM}, we proposed a simple Hamiltonian for the dynamics of the matter-geometry interaction.
If the bosons are not interacting, their total Hamiltonian is trivial, $
\hat H_v = \sum_{i=1}^N \hat H_i =- \sum_i \mu \hat b^\dagger_i \hat b_i.  
$
An interesting interaction term is hopping, the physical process in which a quantum $i$ is destroyed  at $i$ and one  is created at $j$.  We will require that hopping is possible only if there is an {\em on} edge between $i$ and $j$. Such dynamics is described by a Hamiltonian of the form
\begin{equation}\label{hhop}
\hat H_{hop} = -E_{hop}\sum_{(i,j)}\hat P_{ij}\otimes  (\hat b^\dagger_i \hat b_j +\hat b_i \hat b^\dagger_j),
\end{equation}
where
$
\hat P_{ij}=|1\rangle\langle 1|_{(i,j)}
$
is the projector on the  edge $(i,j)$ being in the {\em on} state.
This projector is important, it means that  it is the
dynamics of the particles described by $\hat H_{hop}$ that gives to the link degrees of freedom  the meaning of 
geometry: the state of the graph determines where the matter is allowed to go.

\begin{figure}
\centering
\includegraphics[scale=0.25]{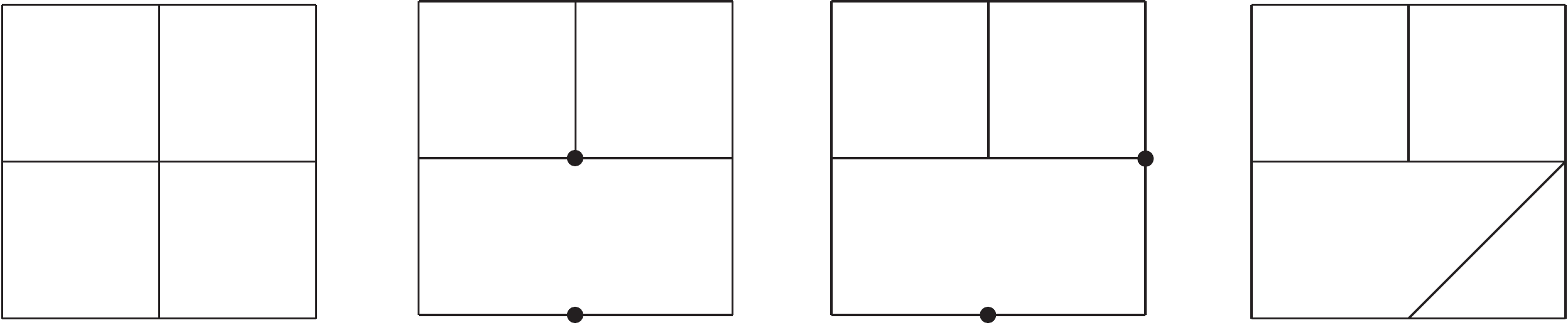} 
\caption{Graph-matter dynamics: A link is is exchanged for two particles at its vertices; the particles hop on existing links; two particles are destroyed and a link is created between the corresponding vertices. }
\label{changinggraph}
\end{figure}

In the spirit of ``geometry tells matter where to go and matter tells geometry how to curve'', we need graph and matter to interact.  To avoid interpretational problems, we also need the interaction to be unitary.  The simplest unitary exchange term is 
\begin{equation}
\hat H_{ex} =k \sum_{(i,j)}  |0\rangle\langle 1|_{(i,j)}\otimes ( \hat b^\dagger_i
\hat b^\dagger_j)^R +|1\rangle\langle 0|_{(i,j)}\otimes (\hat b_i \hat b_j)^R. 
\end{equation}
This destroys an edge $(i,j)$ and create $R$ quanta at $i$ and $R$ quanta at $j$, or, vice-versa, destroys $R$ quanta 
at $i$ and $R$ quanta at $j$ to convert them into an edge.  An example is shown in Figure \ref{changinggraph}.
Of course, we need dynamics also for the graph degrees of freedom alone. The simplest choice is simply 
to assign some energy to every edge,
$
 \hat H_{link} = -U\sum_{(i,j)} \sigma^z_{(i,j)}. $

This final step brings us to the total Hamiltonian for the model proposed in \cite{HMLCSM}:
\begin{equation}\label{h}
\hat H = \hat H_{link}+ \hat H_v + \hat H_{ex} + \hat H_{hop}.
\end{equation}
 It is possible to design such systems in the laboratory. For instance, one can use arrays 
of Josephson junctions whose interaction is mediated by a quantum dot with two levels.

This is a peculiar system in that who interacts with whom is a quantum degree of freedom, but otherwise it is an extremely simple system.  Does it lead to any interesting behavior?  Yes, more than one would expect, as we will see next.  

\subsection{Calculating the speed of light as propagation of information}

$\hat H_{hop}$   tells us that it takes a finite amount of time to go from $i$ to $j$. If the graph was 
 a chain,  it would take a finite amount of time (modulo exponential decaying terms) for a particle to go from one end 
of the chain to another. This results to a ``spacetime'' picture (the evolution of the adjacency graph in time) with 
a finite lightcone structure.
We can calculate this speed of light from the speed with which information propagates using methods from quantum information theory.  
From a local Hamiltonian, that is, a Hamiltonian that is the product of local terms, 
$
H=\sum_{\langle ij\rangle}h_{ij}, 
$
we can define the {\em Lieb-Robinson speed of information propagation} \cite{LR} as follows.  Consider two points $P$ and $Q$ on a lattice,  distance $d_{PQ}$  apart.  A disturbance at $P$ is felt at $Q$ a time $t$ later with strength  $\|[O_P(0),O_Q(t)]\|$, where $O_P(0)$ and $O_Q(t)$ are operators at $P$ at time $0$ and $Q$ at time $t$ respectively.  It is shown in \cite{LR} that this signal strength is bound by 
\begin{equation}
\|[O_P(0),O_Q(t)]\|\leq
2\|O_P\| \|O_Q\|\sum_n\frac{\left(2|t|h_{\rm max}\right)^2}{n!}N_{PQ}(n),
\end{equation}
where $h_{\rm max}$ is the maximum coupling strength in the Hamiltonian and $N_{PQ}$ is the number of paths of length $n$ in the lattice that connect $P$ and $Q$.  This can be rewritten as 
\begin{equation}
\|[O_P(0),O_Q(t)]\|\leq
2\|O_P\| \|O_Q\|
C\exp\left[-a\left(d_{PQ}-vt\right)\right].
\end{equation}
Saturating the above inequality defines the maximum speed $v$ of information propagation in this system.  Combining the two bounds allows us to calculate this speed in terms of the couplings in the Hamiltonian and the connectivity of the lattice.  

In \cite{HMPS}, we tested that $v$ above can be the speed of light, by showing that in string net condensation, a spin system whose emergent excitations are photons \cite{Wen}, $v$ agrees with the speed of light in the emergent Maxwell equations\footnote{
Note, however, that this derivation does not assure us that this maximum speed is {\em universal} for all species of matter.  
}. This is an interesting result as it allows us to reconcile emergent finite light cones and non-relativistic quantum mechanics.  The underlying physics is, of course, quantum mechanics, but, within the bounds defined above,  the system appears to have a finite light cone.  A signal is possible outside the light cone, but it is exponentially suppressed.  In fact, recent results show that in the continuum limit the finite light cone becomes exact as that signal vanishes in the continuum limit.  Further work in \cite{HMPSS2} and results in \cite{eisert} indicate that the emergence of a Minkowski metric is a behavior that can be extended to infinite-dimensional systems, i.e.,  the physics we are studying is not limited to spin systems.

Note that this speed $v$ increases with the number $N_{PQ}$ of paths connecting $P$ and $Q$, and therefore it is a function of the connectivity of the lattice.  Higher connectivity (vertex degree) means higher speed of light.  This is used in what follows to derive the effective curved geometry matter sees.

\subsection{Analogue black holes}

\begin{figure}
\centering
\includegraphics[scale=0.4]{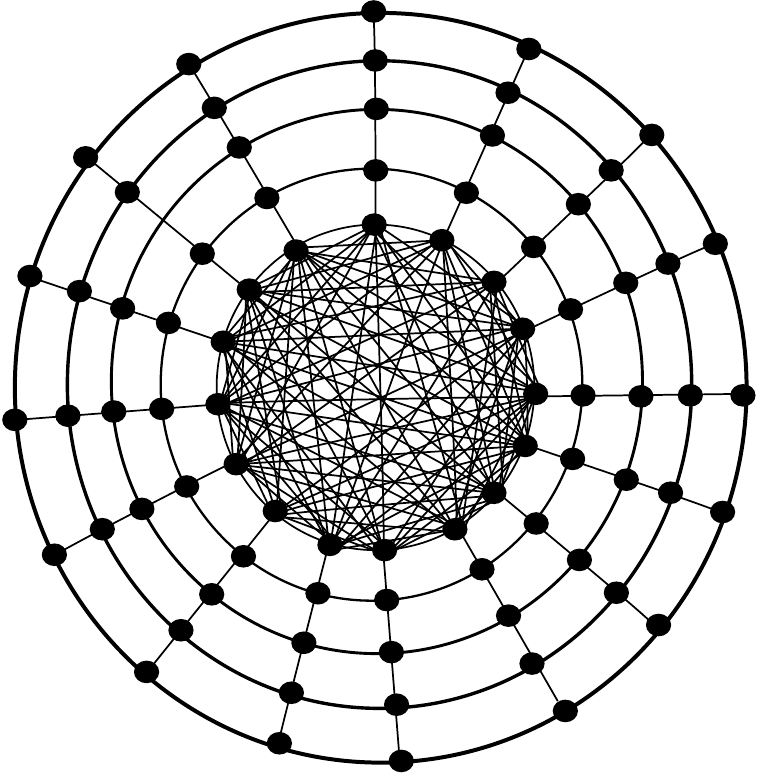} 
\caption{Toy black hole configuration.}
\label{2dkn}
\end{figure}

One of the features of the above Hamiltonian acting on states which are not degree-regular graphs, observed in \cite{HMLCSM}, was the 
trapping of bosons into regions of high degree (see Fig.\ \ref{2dkn}). 
The basic idea is the following: consider two sets of nodes, $A$ and $B$, separated by a set of points $C$ on the 
boundary, and let the vertices in $A$ be of much higher degree than the vertices in $B$, $d_A\gg d_B$ . If the number of edges departing from the set $C$ and going to the set
$A$ is much higher than the number of edges going from $C$ to $B$, then the hopping particles have a high probability 
of being ``trapped'' in the region $A$.

A way to see the trapping is to study the Lieb-Robinson  speed of particle propagation.  Since the speed of propagation of 
particle probability is degree dependent, in the two regions we have two different 
speeds.  Their ratio depends uniquely on 
the ratio of the degree of the two regions. Then, using the laws of optics, probability is reflected 
at the boundary due to Snell's law:  for $\frac{d_B}{d_A}\sim\frac{1}{N}\rightarrow 0$, and the region $A$ acts as a trap. 

This heuristic argument can be made precise. In  \cite{CHMR}, it was shown that 
matter propagating on the graph state shown in Fig.\ \ref{2dkn} has a unique ground state which is protected by a gap which increases 
linearly with the size $N$ of the completely connected region. 
That is, 
high connectivity configurations are spin-system analogues of trapped surfaces.

\subsection{Geometry: what the matter sees}

While it is possible to assign a metric to the graph itself, for the purposes of an analogue model for quantum gravity, the relevant geometry is the one the metric sees.  Clearly, that metric can be affected both by the graph state and by the matter Hamiltonian.  

Solving this problem is not easy.  Our model may be simple conceptually, but it is a Hubbard model on a dynamical irregular lattice and, as is well-known, the Hubbard model beyond 1d quickly becomes intractable.  Luckily, it turns out that a large and interesting sector of our model can be reduced to an effective 1d  Hubbard model with modified couplings. 

 In order to do this, we restrict the time-dependent Schr\"odinger equation to the manifold formed by the classical states, that is, single-particle states with a well-defined but unknown position. The equation of motion obtained corresponds to the equation of propagation of light in inhomogeneous media, similarly to black hole analogue systems. Once we have such a wave equation, we can extract the corresponding metric.
This is a one-dimensional Hubbard model on a lattice with variable vertex degree and multiple edges between the same two vertices. The probability density for the matter obeys a (discrete) differential equation closed in the classical regime. This is a wave equation in which the vertex degree is related to the local speed of propagation of probability. This allows an interpretation of the probability density of particles similar to that in analogue gravity systems: {\em the matter  sees a curved spacetime}. 

\begin{figure}
\centering
\includegraphics[scale=0.8]{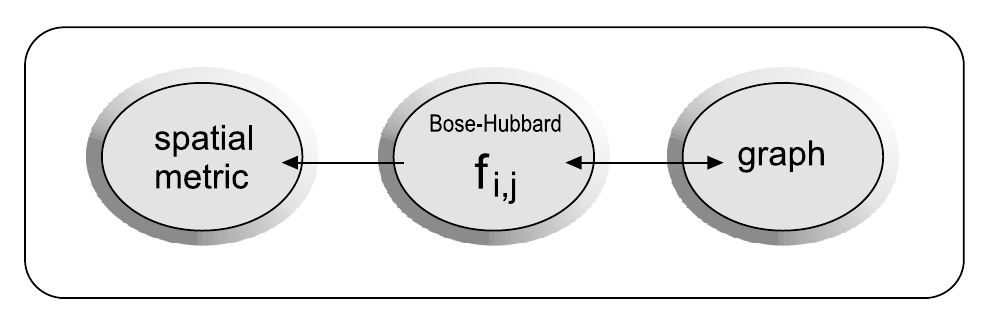}
\caption{The scheme representing how the graph relates the hopping energies $f_{i,j}$ of the Bose-Hubbard model and the emergent metric: the graph modifies the strength of the interaction in the Bose-Hubbard Hamiltonian, which in turn translates into a curved geometry (for the appropriate states).}
\label{fig:scheme} 
\end{figure}

This establishes the desired relation between the connectivity of the graph and the curvature of its continuous analogue geometry.  The overall scheme is illustrated in \ref{fig:scheme}.

\section{Discussion}

We have seen that the idea of the Universe as a Computation is useful because information provides a route to a unification of matter and geometry.  In the above, however, the Computation idea is a tool, not necessarily a fundamental ontology, and our results do not imply or require {\em it from bit}.  It simply helps to talk about information as the primary object.   It can be tempting to interpret this phrase as meaning that {\em bit} pre-exists {\em it}.    However, I must admit that I do not understand what we can mean by information as decoupled from its physical realization.  I prefer to simply take advantage of the fact that it allows us to study a system without having to specify the details of unknown physics.  

In summary, it is {\em possible} to formulate Planck scale physics as a quantum information processing system, as we demonstrated in previous work (Quantum Causal Histories \cite{QCH}).
It is  also {\em useful} to formulate Planck scale physics as a quantum information processing system because:
1. Quantum information provides an unambiguous description of physics before geometry.  
2. It is suitable for emergence problems, just as classical information is useful for statistical physics.
3. It provides a new toolbox well-adapted to background independent problems and, in particular, it lets us import methods from statistical physics to a background independent context.

 Information is useful precisely because it allows us to study the behavior of a system without committing to a particular ontology, necessary when the ontology is ambiguous, as is the case in emergence approaches to quantum gravity.

Even though I am not a believer in the full-blown Universe as a Computation philosophy, it can be fun to explore some of the questions that it raises from the rather concrete viewpoint described above.  We will end with a sample of such questions.

{\em Information as unification.} The old version of unification is the picture of group theory and symmetry breaking and the convergence of fundamental couplings.  While this is now outdated, some level of unification is necessary in quantum gravity to allow for quantum matter to interact with dynamical spacetime, as the language clash between differentiable manifolds  and quantum fields on a fixed background has long been an obstacle to quantum gravity.  The idea that information underlies everything allows a new path: express both gravity and matter in information theoretic terms.  Quantum graphity models are a first step in that direction. It is a long way to go but we are catching a promising glimpse of a novel form of unification.  

{\em Why is the universe so stable?} If  the universe is a computer program, how come it doesn't crash, or at least it hasn't crashed yet?  This sounds like a joke, but it is a relevant question in cosmology: is our universe a stable attractor, and if so why? It is interesting to look for potential commonalities between mechanisms for stability in computers and in physics.  In computers, stability comes from some kind of built-in redundancy that provides error correction.  In physics, certain symmetries can be seen as a kind of error correction.  Elsewhere, we noted that 
   the notion of decoherence-free subsystems used in quantum computing to protect against noise and errors is very similar to the notion of conserved quantities, something we used in 
  \cite{braids}  to find a large class of conserved quantities in Loop Quantum Gravity. I believe these results are only just scratching the surface. 

{\em How are the physics laws/computer program selected?}  Why our universe is what it is is a perennial problem in quantum gravity and cosmology.  In the Universe as a Computation scenario it directly translates to the question of how the Program is selected, and this new viewpoint brings new possibilities.
There are four commonly given answers: 1. Anthropic arguments: by construction, the universe we observe has to be compatible with the conscious life that observes it, hence it is unremarkable that the fundamental constants happen to fall within the narrow range that allows life.  
This is currently a very popular idea, supported by logic and possibly inflation and string theory, but also widely criticized as unscientific and non-explanatory.
2. Our laws have evolved through the history of the universe.  This generally leads to meta-laws selecting the laws and a resulting circular argument.  
3. Multiverse: our universe is one of many physically realized universes.  The many can be arranged in various ways which have been thoroughly classified by Tegmark \cite{Tegmark}.  
Unlike the anthropic argument, this scenario is wider, and, in some forms, in principle testable.  But there is a huge proliferation of potential universes, not just those we can generate by allowing the fundamental constants to take other values (the usual multiverse), but also all possible laws or programs.  This is an instantiation of Tegmark's multiverse. 
4. Ideas of self-organized criticality (SOC): our universe is a stable  attractor.
One may think that this should be the most promising direction, however, such ideas have hardly been explored.  To a great extend, there is a serious technical obstacle.  SOC is typically observed in non-equilibrium systems, while all of fundamental physics uses equilibrium quantum field theory.  Properly introducing SOC ideas in cosmology requires a departure from the standard framework.  
Since many of the results in this area are already expressed in algorithmic terms,  a description of the Universe as a computation can make it easier to introduce SOC ideas to a (quantum) cosmological setting.  It will, of course, be necessary to study quantum systems that exhibit SOC  first.   This is a fascinating long-term direction for this kind of work.  

\section*{Acknowledgements}
This work is supported by the Alexander von Humboldt and the Templeton Foundations.  Research at Perimeter Institute is supported by the Government of Canada
through Industry Canada and by the Province of Ontario
through the Ministry of Research \& Innovation.

\end{document}